\documentclass[colorlinks=true, linkcolor=blue, citecolor=blue, filecolor=blue, urlcolor=blue]{aa}
\usepackage{verbatim}

\usepackage{graphicx}
\usepackage{txfonts}
\newcommand{\aref}[1]{\hyperref[#1]{Appendix~\ref{#1}}}

\usepackage{orcidlink}

\usepackage{url}
\begin{document}

\title{Radiation magnetohydrodynamics simulations of Population III star formation during the Epoch of Reionization}

\author{Lisanne van Veenen$^{\orcidlink{0009-0005-8612-4059}}$\inst{1},
Piyush Sharda$^{\orcidlink{0000-0003-3347-7094}}$\inst{1},
Serena Viti$^{\orcidlink{0000-0001-8504-8844}}$\inst{1,2,3}
\and
Shyam H. Menon$^{\orcidlink{0000-0001-5944-291X}}$\inst{4,5}
}

\institute{Leiden Observatory, Leiden University, P.O. Box 9513, 2300 RA Leiden, The Netherlands\\
\email{vanveenen@strw.leidenuniv.nl; sharda@strw.leidenuniv.nl}
\and
Transdisciplinary Research Area (TRA), Argelander-Institut für Astronomie, University of Bonn, 53113 Bonn, Germany
\and
Department of Physics \& Astronomy, University College London, London WC1E 6BT, United Kingdom
\and
Center for Computational Astrophysics, Flatiron Institute, 162 5th Avenue, New York, NY 10010, USA
\and
Department of Physics and Astronomy, Rutgers University, 136 Frelinghuysen Road, Piscataway, NJ 08854, USA
 }

\date{Received November 14, 2025; accepted }

% \abstract{}{}{}{}{} 
% 5 {} token are mandatory

\abstract{Cosmological simulations find that pockets of star-forming gas could remain pristine up until the Epoch of Reionization (EoR) due to the inhomogeneous nature of metal mixing and enrichment in the early Universe. Such pristine clouds could have formed Population III stars, which could have distinct properties compared to their very high redshift ($z \geq 20$) counterparts. We investigate how Population III stars form and grow during the EoR, and whether the resulting mass distribution varies with environment or across cosmic time. We perform high-resolution ($7.5\,\rm{au}$) radiation-magnetohydrodynamics simulations of identical primordial clouds exposed to the CMB appropriate for $z=30$ and $z=6$, respectively, as part of the POPSICLE project. We also run a simulation at $z=6$ with a strong external Lyman-Werner (LW) background, to span across radiative environments which could host metal-free clumps during the EoR. In the limit of no external LW radiation, we find that while the evolution of the most massive star ($M_{\star} \approx 70\,\rm{M_{\odot}}$) is almost identical between $z=30$ and $z=6$, the latter exhibits less fragmentation, leading to a smaller cluster of stars with a higher median stellar mass. In the limit of high external LW radiation, we see vigorous accretion and high star formation efficiencies, leading to the formation of very massive ($M_{\star} > 100\,\rm{M_{\odot}}$) stars. Our results suggest that Population III IMF could vary with redshift simply due to the CMB, independent of the environment. We find that less massive and more compact Pop III star clusters could form during the EoR as compared to $z \geq 20$, with the formation of very massive and supermassive stars likely in strongly irradiated environments.}


\titlerunning{Simulating Pop III star formation during the EoR}
\authorrunning{van Veenen et al.}

\maketitle

\section{Introduction}
\label{s:intro}
It has been long thought the first generation of stars (Population III) formed from primordial gas around $200-300\,\rm{Myr}$ after the Big Bang \citep[e.g.,][]{2000ApJ...540...39A,abel_seminal_pathway,Bromm_2002,2006ApJ...652....6Y}. The majority of Population III stars are believed to have formed with a top-heavy initial mass function (IMF), although the exact shape of Population III IMF and possible variations therein remain a topic of active debate \citep[see the recent review by][]{klessen_glover}. Low mass ($M_{\star} < 0.8\,\rm{M_{\odot}}$) Population III stars that could have survived today would be rare \citep[][]{2014ApJ...785...73S,2015MNRAS.447.3892H,2021MNRAS.503.6026R} and extremely challenging to identify \citep[e.g.,][]{2015ARA&A..53..631F,where_lowmass_popiii,2018MNRAS.473.5308M}. More than 20 hours of observing time on the James Webb Space Telescope (JWST) has been dedicated to Population III star searches, but so far only tentative evidence for a Population III galaxy or cluster has been reported \citep{2024A&A...687A..67M,fujimoto2025glimpseultrafaintsimeq105}. Detecting light from the era where they dominated the cosmic star formation rate density (SFRD) at $z = 20-30$ \citep{klessen_glover,Venditti_2023} would require instruments capable of detecting objects as faint as 39 $\text{mag}_{AB}$ which could only be achieved using the hypothetical \textit{Ultimately Large Telescope} with a mirror of a diameter of $100\,\rm{m}$ placed on the moon \citep{ULT}. These constraints limit the possibility of directly detecting Pop III stars at an epoch where they are expected to be abundant, compounded by the fact that most of them would have survived for a short time period given their large masses \citep{2002ApJ...567..532H}.
 
Earlier works demonstrated the possibility that Population III stars might have continued to form during the Epoch of Reionization (EoR) at $z = 6-9$ due to the inhomogeneous nature of metal mixing and enrichment in the early Universe (\citealt{2009ApJ...700.1672T,2013MNRAS.428.1857J,2013ApJ...773...19M}; more recently -- \citealt{2020MNRAS.497.2839L}). Recent cosmological simulations with more sophisticated treatments of baryonic physics further support the idea of such a late mode formation of Pop III stars \citep{2016ApJ...823..140X,Venditti_2023,zier2025}. These simulations show that during the EoR, gas clouds situated in low-mass haloes on the outskirts of a (proto-)galaxy at distances up to $\sim 20\,\rm{kpc}$ from the galactic center could remain pristine for a substantial time. These findings open an exciting avenue for targeted searches of Pop III stars that could be possible with the current generation of facilities \citep{2010MNRAS.404.1425J,2023MNRAS.524..351K,2023jwst.prop.3516M,2025arXiv250922776W}. 

While cosmological simulations provide a detailed picture of the large scale environment within which such a late mode formation of Pop III stars is possible, they cannot resolve the formation and growth of individual Population III stars, and hence cannot predict an IMF. Semi-analytical models that track the cosmic evolution of Pop III star formation and the Pop III SFRD \citep[e.g.,][]{2003ApJ...589...35S,2006MNRAS.373..128G,2018MNRAS.475.5246V,2018MNRAS.479.4544M,2024MNRAS.535..516H,2024MNRAS.529..628V,2025arXiv250719581H} suffer from similar issues, and do not track all relevant physical processes that can influence the Pop III IMF. As such, it is not yet known what were the masses and radiative properties of Pop III stars that formed during the EoR, and whether they were any different than those at $z \geq 20$ as expected from small box but very high resolution hydrodynamical simulations of collapsing primordial clouds \citep[][and references therein]{klessen_glover}. As a result, there is a substantial scatter in UV luminosity functions and cosmic SFRDs predicted for an EoR Pop III stellar population, which would have implications for their potential observability \citep[][figures 11 and 12]{fujimoto2025glimpseultrafaintsimeq105}. Making theoretical predictions on the properties of Population III stars during the EoR is essential now that the first direct Population III detection could be within reach thanks to JWST \citep[e.g.,][]{2018ApJ...854...75S,2021ApJ...918....5B,2022MNRAS.513.5134N,2022MNRAS.512.3030V,2023MNRAS.525.5328T,2024MNRAS.533.2727Z,2024ApJ...973L..12V,2025arXiv250520263V,2025ApJ...989L..32R}.  

Motivated by these possibilities, we carry out the first 3D radiation-magnetohydrodynamics (RMHD) simulations of Population III star formation during the EoR. Our goal is to understand the differences, if any, in the masses and properties of Pop III stars at $z=6$ as compared to $z \geq 20$ when both magnetic fields and stellar radiation feedback are taken into account. We present our simulation setup in \autoref{s:setup}, show and discuss the results in the case of no external radiation in \autoref{s:results}, explore how external radiation can impact the mass growth of Pop III stars in \autoref{s:bkg}, and finally conclude in \autoref{s:conclusions}. 

\section{POPSICLE Simulations Suite}
\label{s:setup}
We provide a brief overview of the POPSICLE Project (POP III/II Simulations Including Chemistry, Luminosity and Electromagnetism) setup we use to run the RMHD simulations in this work. For further details on the numerical setup, we refer the reader to \citet{Sharda_2025_popsicle_explanation_notthebigonewithme}. We use the adaptive mesh refinement (AMR) code FLASH \citep{2000ApJS..131..273F,flash}. We use a customized version of the compressible MHD solver developed by \citet{Bouchut2007,Bouchut2010} in our simulations \citep{2009JCoPh.228.8609W,2011JCoPh.230.3331W}. We employ the VETTAM radiation hydrodynamics scheme, which uses a non-local variable Eddington Tensor (VET) closure obtained with a ray-trace solve to close the radiation moment equations \citep{Menon_2022_VETTAM}. We couple VETTAM to the KROME astrochemistry package \citep{grassi_2014} to model non-equilibrium primordial chemistry (including deuterium and three-body H$_2$ formation) with H/H$_2$ photoionization and dissociation rates computed from the radiation solution (Menon et al. in prep). We evolve the radiative properties of the protostars as a function of their mass and accretion rates using the \texttt{GENEVA} Pop III protostellar models \citep{haemmerle_2018}. The protostars are represented as sink particles, following the criteria outlined by \citet{Federrath_2010}. We do not allow sink particle mergers unless explicitly specified (cf. \autoref{s:bkg_merge}).

The radiative output from Population III protostars becomes significant for accretion rates below $\dot M_{\rm{crit}} = 0.01\,\rm{M_\odot\,yr^{-1}}$ \citep[e.g.,][]{Omukai_2003_popiii_acc,2009ApJ...703.1810H,haemmerle_2018}. We model the ionization of H and H$_2$ by extreme-UV (EUV) photons, with two bands corresponding to energies $h\nu > 13.6\,\rm{eV}$ for H and $h\nu > 15.2\,\rm{eV}$ for H$_2$ ionization respectively \citep[e.g.,][]{2015MNRAS.454..380B,Sharda_2025_popsicle_explanation_notthebigonewithme}. We also include dissociation of H$_2$ by far-UV (FUV) photons in the Lyman-Werner (LW) band ($11.2\,\mathrm{eV} < h\nu < 13.6\,\mathrm{eV}$), which can have significant effects on the thermodynamics of dense gas. This LW radiation can originate either in-situ from the protostar(s), or from an external UV background formed by nearby galaxies. Lastly, we incorporate not only self-shielding of H$_2$ but also the often ignored cross-shielding of H$_2$ by H \citep{Draine_Bertoldi_1996,2011MNRAS.412.2603W,wolcott_green_shielding_2011}, which can have a significant impact on the protostellar mass growth and ultimate fate of Pop III stars (Chen et al. in prep). 

To put the context of Pop III star formation during the EoR in the broader literature of Pop III studies, we keep the initial conditions identical to the simulations at $z=30$ by \citet{sharda2025} except that we set $z = 6$ in our work. This essentially means that we are testing the impact of the cosmic microwave background (CMB) radiation on primordial star formation. While the CMB does not matter for star-forming gas at $z<5$, it can appreciably modify gas thermodynamics and the IMF at high-redshifts, regardless of the metallicity \citep[e.g.,][]{2022MNRAS.514.4639C,2023MNRAS.519..688B,2025MNRAS.537..752B}. Between $z=6$ and $z=30$, the CMB temperature, $T_{\rm{CMB}}$, decreases by $\Delta T_{\rm{CMB}} \approx 66\,\rm{K}$, which is a substantial difference even for primordial star-forming gas \citep[e.g.,][]{Galli_1998,omukai_2005}. 
 
The initial cloud mass and radius are $1000\,\rm{M_{\odot}}$ and $1\,\rm{pc}$, respectively. We initially stir turbulence by implementing a randomly generated trans-sonic velocity field (Mach unity). We set an initial rms magnetic field strength of $28\,\mu\rm{G}$, emulating the upper floor to amplification by small-scale dynamo in primordial conditions predicted by models \citep[e.g.,][]{2010ApJ...721L.134S,2011_federrath,schober_2012,schober_2015}, and also seen in simulations \citep[e.g.,][]{Turk_2012,2021MNRAS.503.2014S,2024A&A...684A.195D}. This field strength is equal to 10 percent of the initial turbulent kinetic energy, and the initial field geometry is completely random without any preferred orientation, as would be expected for a small-scale turbulent dynamo. We initialize the cloud with solid-body rotation, such that its rotational energy is 3 per cent of its gravitational energy \citep[e.g.,][]{Bromm_2002, 2006ApJ...652....6Y}. We refine with 64 cells per Jeans length, significantly higher than hydrodynamics simulations that include radiation feedback, to correctly capture the effects of shocks on gas chemistry and thermodynamics \citep[][appendix A]{2021MNRAS.503.2014S}. We allow 10 levels of grid refinement based on the Jeans length, which leads to a maximum spatial resolution of $\Delta x = 7.5\,\rm{au}$, yielding the density threshold for sink particle formation to be $10^{13}\,\rm{cm^{-3}}$. Following \citet{2011IAUS..270..425F}, we set the accretion zone of the sink particle to be $2.5\Delta x$.

Another important distinction between Pop III star formation sites at $z=6$ and $z \geq 20$ is that the former is likely impinged by external radiation from nearby galaxies due to the lower opacity of the intergalactic medium (IGM) to UV radiation at $z=6$ \citep[e.g.,][]{2020A&A...641A...6P,2022MNRAS.514...55B}, or because these Pop III clouds could be present in the vicinity of Pop II starbursts \citep[e.g.,][]{2013MNRAS.428.1857J,2017ApJ...844..111C,Venditti_2023}. Thus, by neglecting external radiation, our setup represents a conservative lower bound on pristine star-forming environments likely present during the EoR. We will discuss the impact of a (strong) external radiation field later in \autoref{s:bkg}.

\section{Pop III star formation at $z=30$ versus $z=6$}
\label{s:results}
In this section, we present and discuss the results from the two simulations (without external radiation) at $z=30$ and $z=6$, respectively, and highlight how differences in the CMB can impact Pop III star formation.

\subsection{Stellar evolution}
\label{s:results_stars}
\begin{figure}
\centering
\includegraphics[width=\columnwidth]{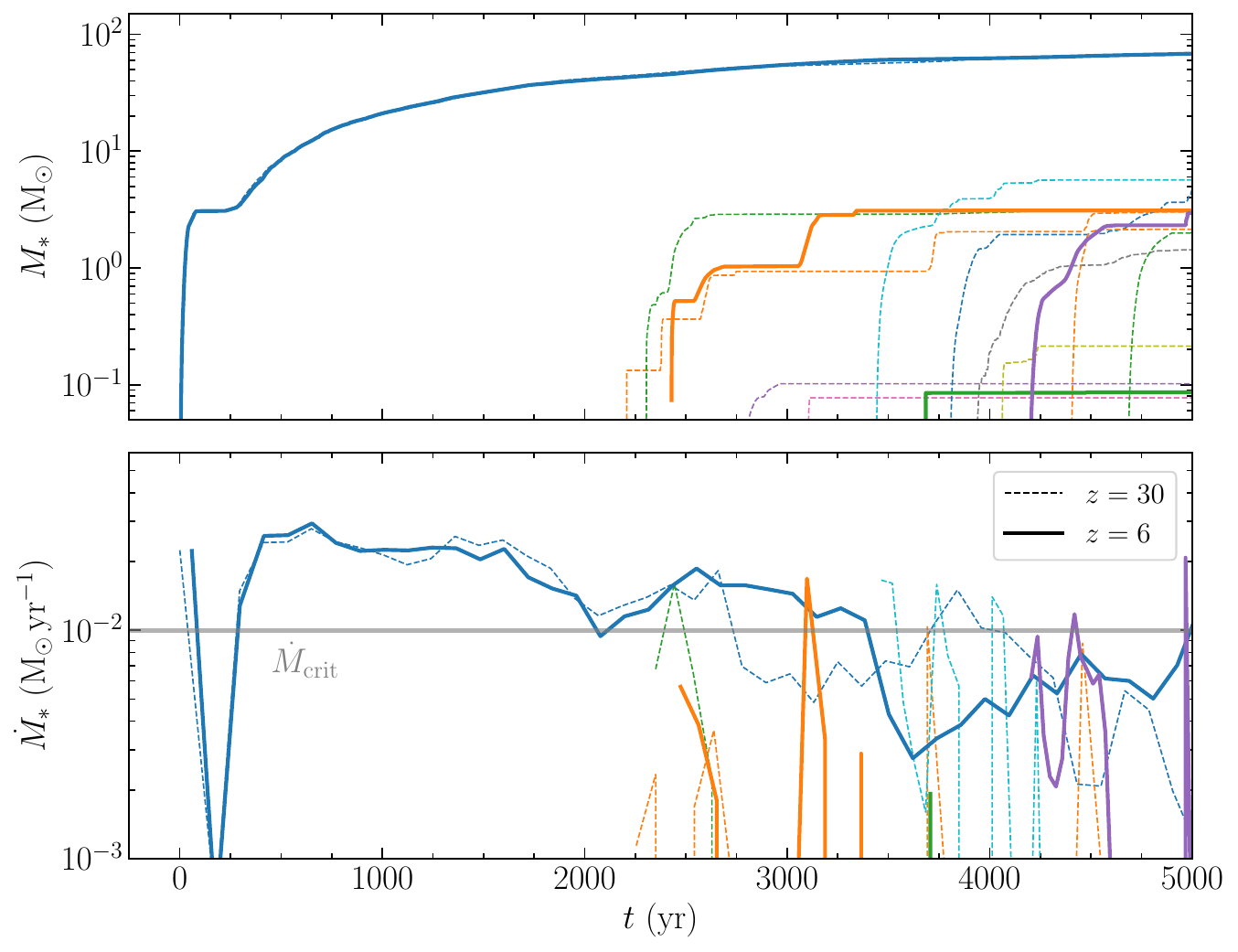}
\caption{Evolution of the individual stellar masses of stars (top), and the mass accretion rate (bottom) for the $z = 6$ (solid) and $z=30$ (dashed) simulations. $\dot M_{\rm{crit}}$ denotes the critical accretion rate below which radiation from Population III stars becomes significant.} Note that the accretion rates of only the four most massive stars are shown at $z=30$ for clarity.
\label{fig:nobg_sinkmass}
\end{figure}

In the time period we simulate, fragmentation leads to the formation of several stars in both the $z=30$ and $z=6$ simulations. In both the simulations, the growth of the oldest and the most massive star is almost identical, as we read from \autoref{fig:nobg_sinkmass} which shows the time evolution of the stellar masses and gas accretion rates. Specifically, the first stars grow to $68\,\rm{M}_\odot$ at both the redshifts. It is interesting to note that these stars do not become very massive ($M_{\star} > 100\,\rm{M_{\odot}}$), which is possible in the absence of magnetic fields as the protostars would have been able to accrete more rapidly \citep{sharda2025}, before stellar radiation feedback can cut off accretion at later times \citep{2011Sci...334.1250H}. The zero age main sequence (ZAMS) mass would not be too different given that accretion onto these objects has declined to rates much below $0.01\,\rm{M_{\odot}\,yr^{-1}}$.

However, contrary to prior expectations, the number and masses of the stars that form next show some surprising differences. We find that 12 and 4 stars form in the $z = 30$ and $z = 6$ simulations, respectively. Intuitively, a lower $T_{\rm{CMB}}$ would lead to lower Jeans masses and therefore a higher degree of fragmentation, but we observe less fragmentation at $z=6$. The accretion rates of the secondary stars manifest in brief, intense bursts at both redshifts. This can be explained by their orbit around the center where they encounter dense pockets of gas leading to brief episodes of intense accretion. As a consequence of these episodic bursts, the stellar masses evolve with a few sudden increases after which they plateau. The stellar masses of the secondary stars span from $0.1 - 8\,\rm{M_{\odot}}$, producing brown dwarfs to red giant stars. At $z=6$, the four stars are locked in a quadruple system \citep[following the multiplicity criteria outlined in][]{bate_2009_merger}. At $z=30$, we instead find both a quadruple and a binary system, while the remaining stars are isolated. The least massive stars stop accreting by the end of the simulations, due to a combination of dynamical effects (e.g., ejection from gas rich regions) and fragmentation-induced starvation \citep[e.g.,][]{2010ApJ...725..134P,2022MNRAS.510.4019P}. The median stellar mass at $z=6$ is larger than that at $z=30$ by 50 percent, and the total mass in stars is smaller by 16 percent, which is not surprising given less fragmentation in the former case.

Given that the Kelvin-Helmholtz time tends to strongly increase towards lower masses for Population III stars \citep[e.g.,][]{schaerer_god}, the intermediate and low mass stars will have protostellar contraction phases that far surpass the time period we simulate in this work. Therefore, our simulations only span the earliest phases of protostellar evolution for these stars. On the other hand, Population III stars with very high mass tend to reach the main sequence very quickly, with estimates at $\lesssim 10^4\,\rm{yr}$ for stars with $M_{\star} \gtrsim 50\,\rm{M}_\odot$ \citep[e.g.,][]{nandal_2023,2023AJ....165....2L}. Thus, our simulations are unlikely to miss a significant portion of the protostellar evolution of massive Pop III stars.

\subsection{Chemistry and thermodynamics}
\label{s:results_chem}
\begin{figure}
\centering
\includegraphics[width=\columnwidth]{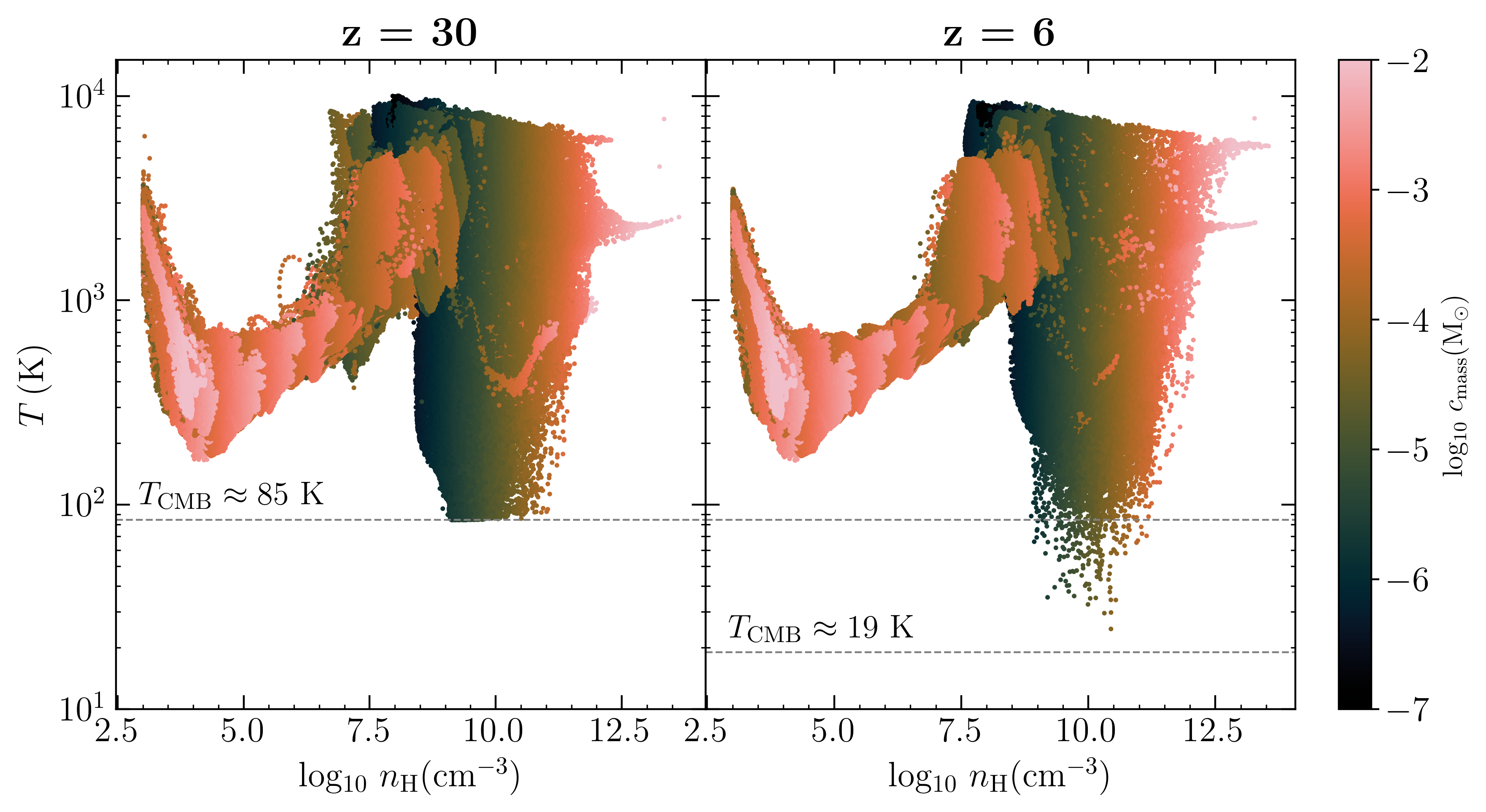}
\caption{Phase diagrams of the star-forming core in the $z=30$ and $z=6$ simulations $5000\,\rm{yr}$ since first star formation, showing the gas temperature plotted against the number density, color coded by the cell mass. The CMB temperature at $z=30$ and at $z=6$ is indicated with gray, dashed lines.}
\label{fig:phaseplot_nobg}
\end{figure}

We show the gas phase distribution at $t = 5000\,\rm{yr}$ in \autoref{fig:phaseplot_nobg}. The two simulations show a globally similar thermodynamic profile, except that in the $z=6$ run, certain pockets of the accretion disk are able to cool down to temperatures below the CMB temperature at $z=30$. Given the controlled nature of these simulations where only the redshift has been changed, it is likely that the different mass and density distributions of protostars between the two runs are correlated with the decreased $T_{\rm{CMB}}$ at $z=6$. At lower $T_{\rm{CMB}}$, the densest gas has a tendency to build up around the center of mass instead of forming secondary clumps (fragmentation) or filaments that are observed in the case with higher $T_{\rm{CMB}}$.

\begin{figure*}
\centering
\includegraphics[width=\textwidth]{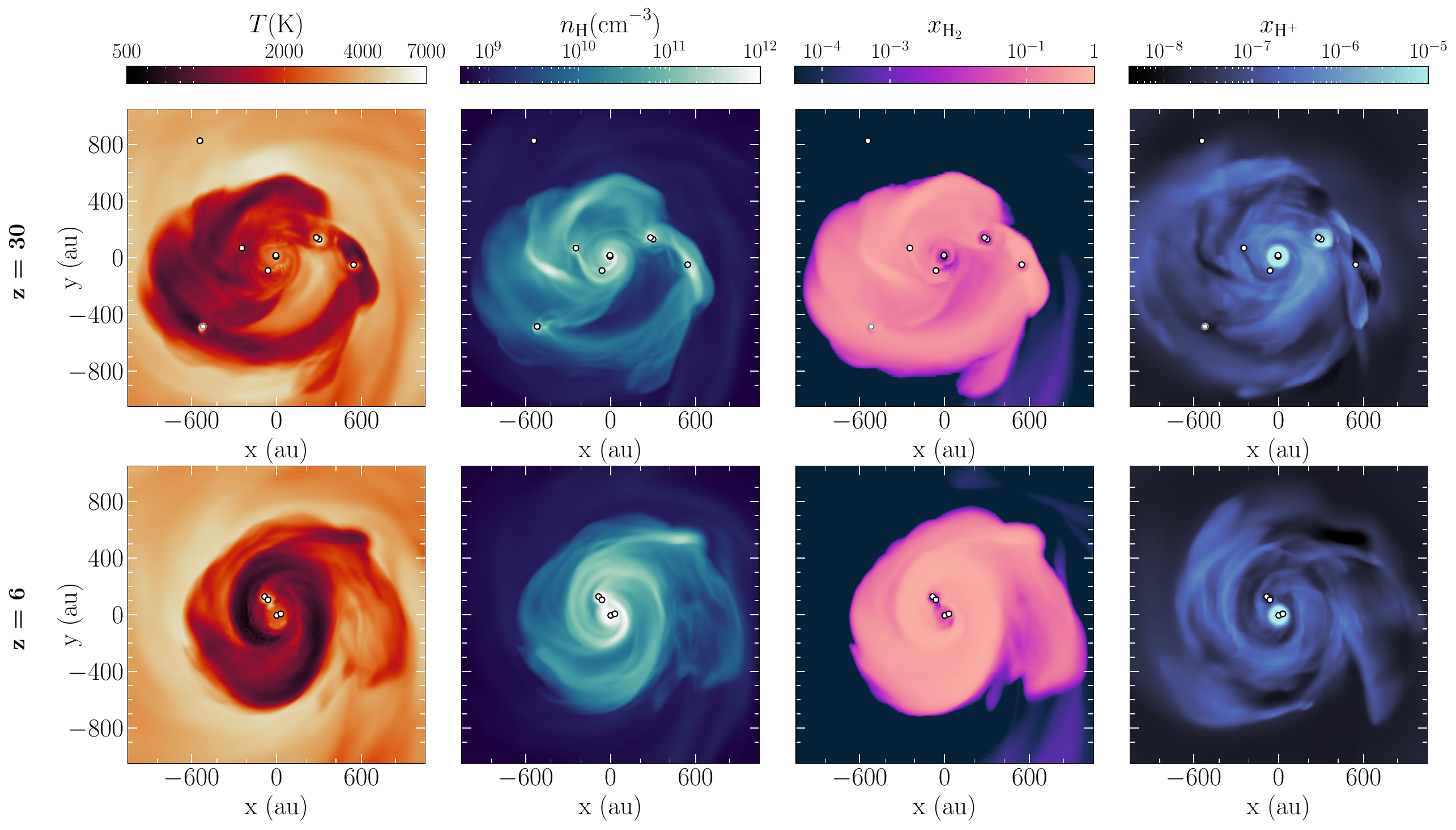}
\caption{Density-weighted projection plots of gas temperature, number density, and $\rm{H}_2$ and $\rm{H}^{+}$ mass fractions $5000\,\rm{yr}$ since the formation of the first star. The projections are oriented along the $\hat{L}$ axis (angular momentum vector of the star forming disk), zoomed in on the inner $0.01\,\rm{pc}$ around the center of mass (COM). The top row shows the $z=30$ simulation and the bottom row shows the run at $z=6$. White dots mark the projected positions of sink particles, which serve as proxies for Pop III stars.}
\label{fig:nobg_overview_basic}
\end{figure*}

\begin{figure}
\centering
\includegraphics[width=\columnwidth]{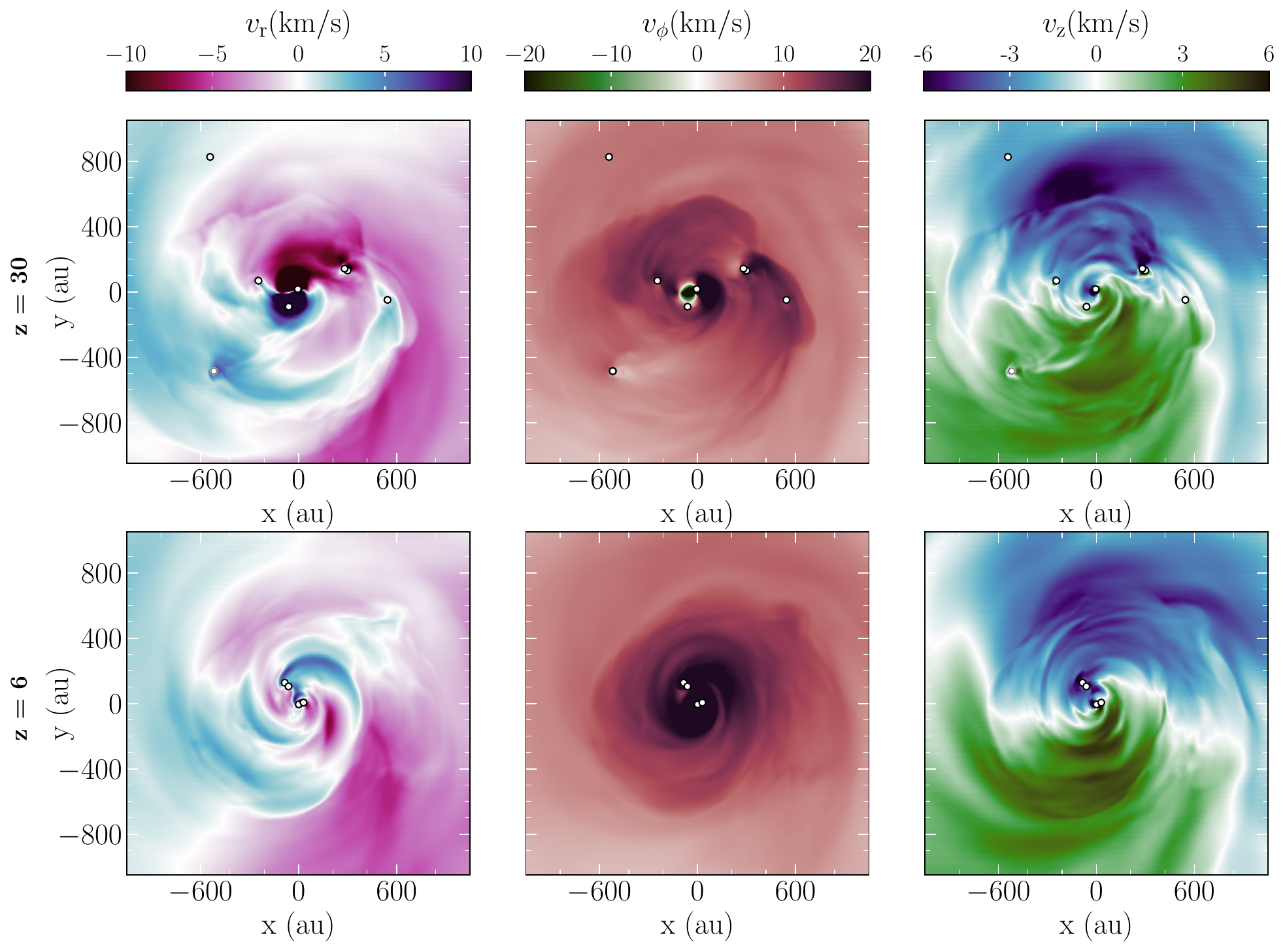}
\caption{Same as \autoref{fig:nobg_overview_basic} but for the radial, azimuthal (rotational), and vertical velocity. Negative radial motions reflect infall of gas towards the COM.}
\label{fig:velocities_nobg}
\end{figure}

We further look into the thermodynamical and chemical properties of the gas, together with the spatial distribution of the stars, in \autoref{fig:nobg_overview_basic}. At $z=6$, a spiral stream of relatively cold ($T \sim 500$ K), dense gas surrounds the center. We do not see this feature at $z=30$, where the central region has larger amounts of hotter, lower density gas. Furthermore, the disk at $z=6$ has evolved to a more compact and tightly wound structure compared to $z=30$, which implies higher rotational velocities and angular momentum. Additionally, at $z=6$ the stars are concentrated in the central, densest region of the gas while at $z=30$ the stars are spread out over much larger distances along individual, denser filaments of the gas. 
The high density gas is fully molecular at $T \sim 500\,\rm{K}$, as we corroborate from the mass fraction of $\rm{H}_2$ we show in \autoref{fig:nobg_overview_basic}.

Two partially ionized \ion{H}{ii} regions are visible in the fourth panel for $z=30$, and one for $z=6$, indicating that the central stars (2 for each \ion{H}{ii} region) have started to emit significant amounts of ionizing radiation as the accretion rates drop, and the star starts to radiate away enough energy to contract to lower radii and raise its temperature \citep[e.g.,][]{schaerer_god,2009ApJ...703.1810H}. 

\subsection{Gas kinematics, magnetic fields and turbulence}
\label{s:results_kinematics}

\begin{figure*}
\centering
\includegraphics[width=\textwidth]{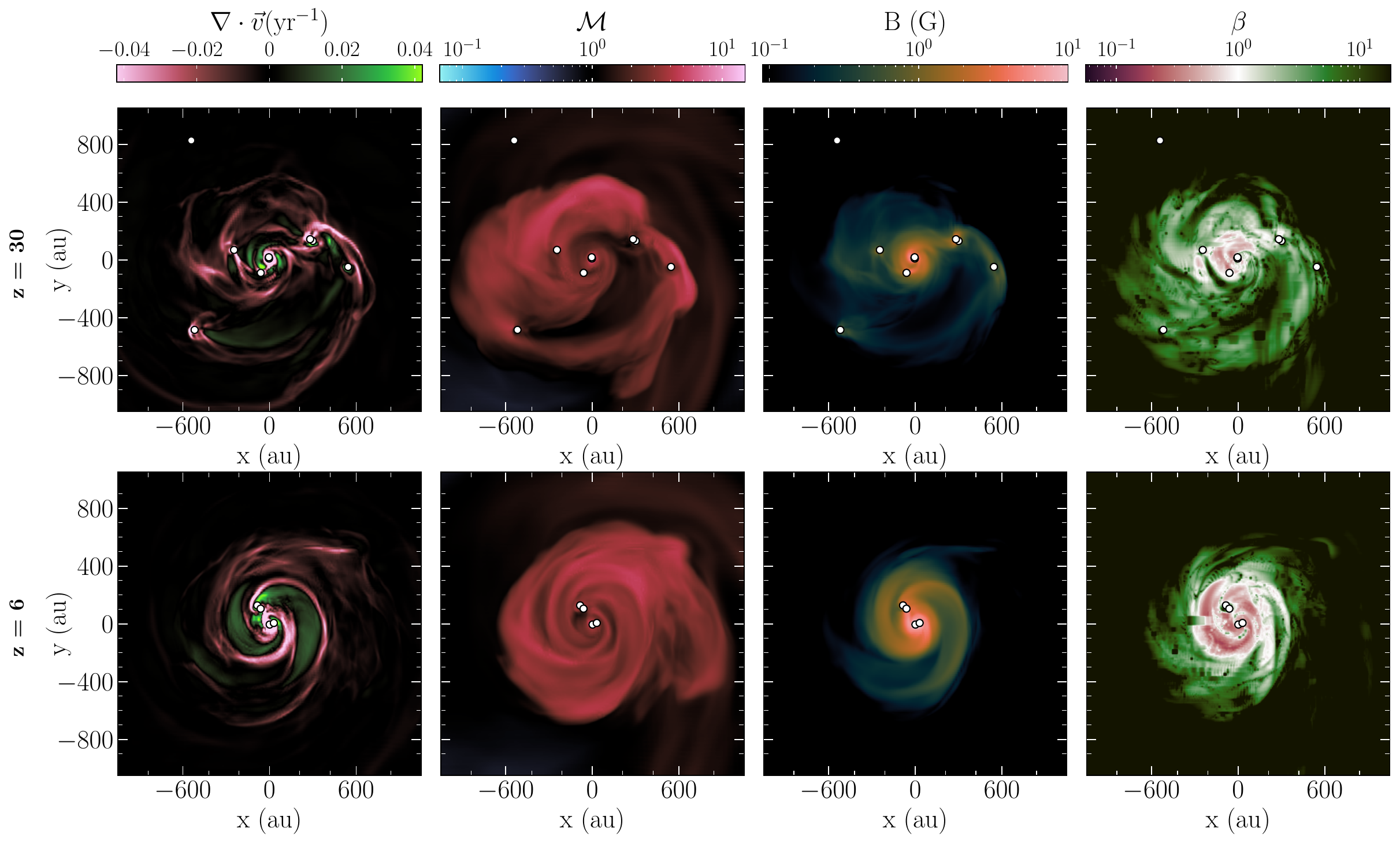}
\caption{Density-weighted projections along the angular momentum vector of the star forming disk, showing how mass inflows are impacted by magnetic fields and turbulence. From left to right, the panels show the divergence of the gas velocity field, turbulent Mach number, magnetic field strength, and the ratio of gas pressure over magnetic field pressure through plasma $\beta$. Negative values of $\nabla \cdot \vec{v}$ indicate net inflow/compression of gas. $\mathcal{M} > 1$ indicates supersonic turbulence and plasma $\beta < 1$ indicates that the magnetic field is dominant as compared to thermal pressure.}
\label{fig:nobg_mag}
\end{figure*}

We show the kinematics of the gas in the disk in \autoref{fig:velocities_nobg}. We find much higher (negative) radial velocities, $v_{\rm{r}}$, at $z=30$ than at $z=6$, which rapidly transports gas from the outer parts of the disk to the inner regions. In the vertical direction, the gas shows very similar behavior between the two redshifts, with sub-dominant motions as compared to those in the radial and azimuthal directions. On the other hand, we find higher rotational velocities, $v_{\phi}$, in the $z=6$ run as compared to $z=30$. At $z=6$, the entire disk remains intact and shows coherent rotation whereas at $z=30$, enhanced fragmentation produces numerous sinks that disrupt the disk and draw mass and angular momentum from the surrounding gas, leading to less coherent rotation. This is further reflected in the divergence field of the gas velocity (see \autoref{fig:nobg_mag}), which depicts how shocks that form within the spiral arms funnel gas towards the central region. At $z=30$, we observed enhanced mass buildup at sites away from the center where others stars have formed. The gas develops supersonic turbulence in both cases, likely driven by gravitational instabilities, with turbulent Mach numbers $\mathcal{M} \sim 10$, as we show in the second column of \autoref{fig:nobg_mag}. 

The third column of \autoref{fig:nobg_mag} shows that the magnetic field strength can reach as high as few $\rm{G}$ throughout the central region. In both cases, the magnetic field peaks at the center around the most massive star due to flux-freezing. The field is stronger near the centre due to larger magnetic pressure as compared to the thermal pressure, leading to plasma $\beta < 1$. However, the field strength mostly remains sub-dominant as compared to turbulence, yielding Alfvén Mach numbers ($\mathcal{M}_{\rm{A}} = \mathcal{M}\sqrt{\beta/2}$) that are larger than unity on average across the disk. Given that the Jeans length $\lambda_{\rm{J}} \propto (1+\beta^{-1})^{1/2}$ \citep[e.g.,][]{2012ApJ...761..156F}, the dominating magnetic field pressure leading to lower $\beta$ values throughout a larger region of the disk at $z=6$ implies larger Jeans lengths and Jeans masses. 
%In the final mass distributions in Table  \ref{tab:masses_nobg} a much smaller population of low mass stars is seen, further adding to this suggestion. 
Therefore, enhanced magnetic activity could also play a role in reducing fragmentation at $z=6$. Although the magnetic field strength is relatively high, we do not expect non-ideal MHD effects to significantly modify the operation of the small-scale dynamo. This expectation is supported by previous theoretical studies \citep{2020MNRAS.496.5528M,nakauchi_mag} and numerical simulations \citep{2023MNRAS.519.3076S}, which find that the effective diffusivities associated with non-ideal MHD processes remain small compared to the amplification driven by the small-scale dynamo during Pop III star formation. However, non-ideal effects may play a more important role on larger scales \citep[][]{2021ApJ...909...37L}, where they can suppress the large-scale dynamo and lead to reduced mean magnetic field strengths \citep[][section 3.4]{2021MNRAS.503.2014S}, potentially hindering the launch of protostellar jets and outflows.

\begin{figure}
\centering
\includegraphics[width=\columnwidth]{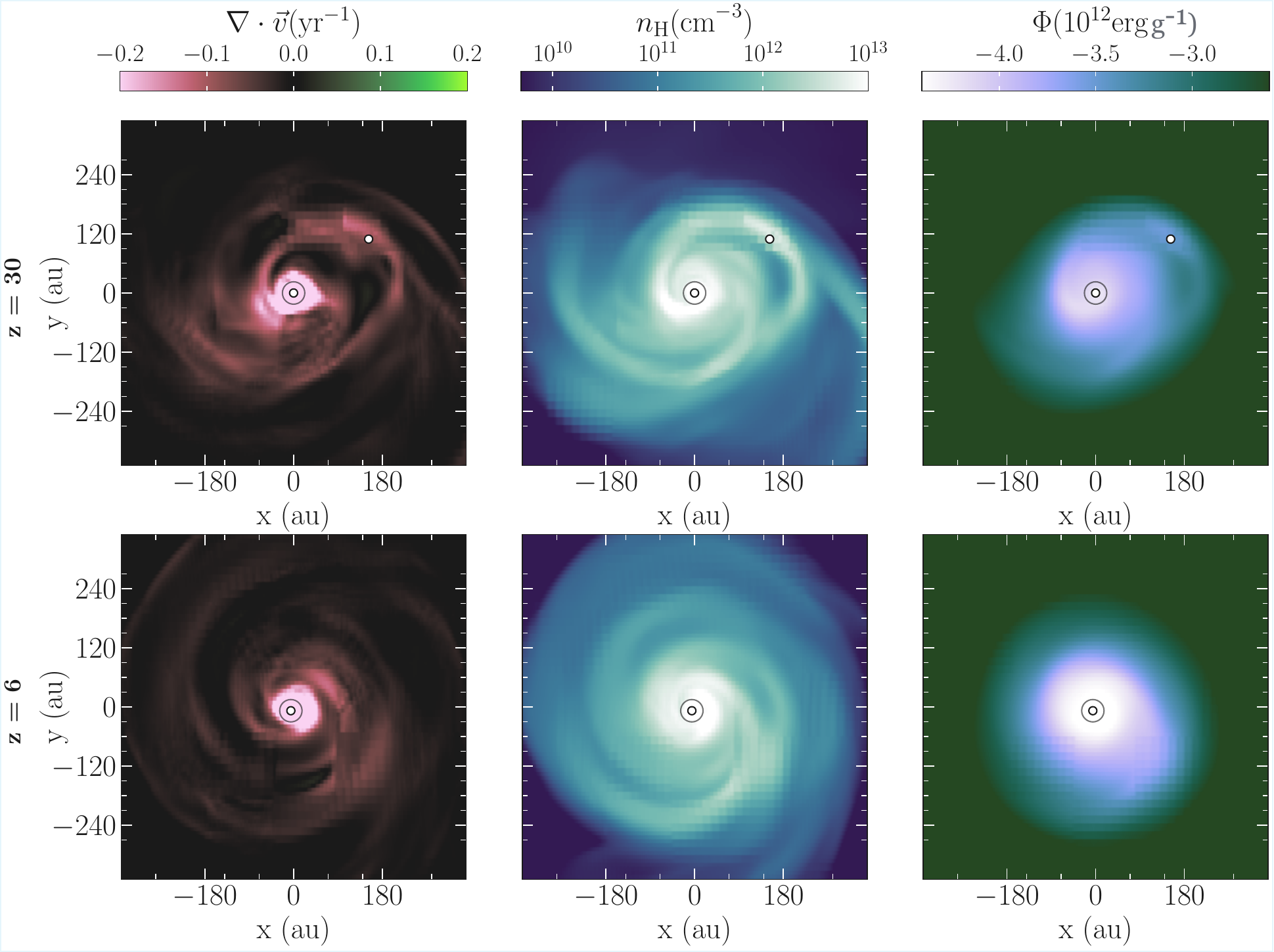}
\caption{Comparison of some of the key criteria for sink particle formation, at $z=30$ (top row) and at $z=6$ (bottom row), right after the formation of the second sink particle at $z=30$. From left to right, the columns show density-weighted projections of the divergence of the gas velocity, number density, and gravitational potential. The projected sink particle positions are indicated using white dots, with rings demarcating the accretion zone of the sink particles.}
\label{fig:sink_criteria_nobg}
\end{figure}

\begin{figure*}
\centering
\includegraphics[width=\textwidth]{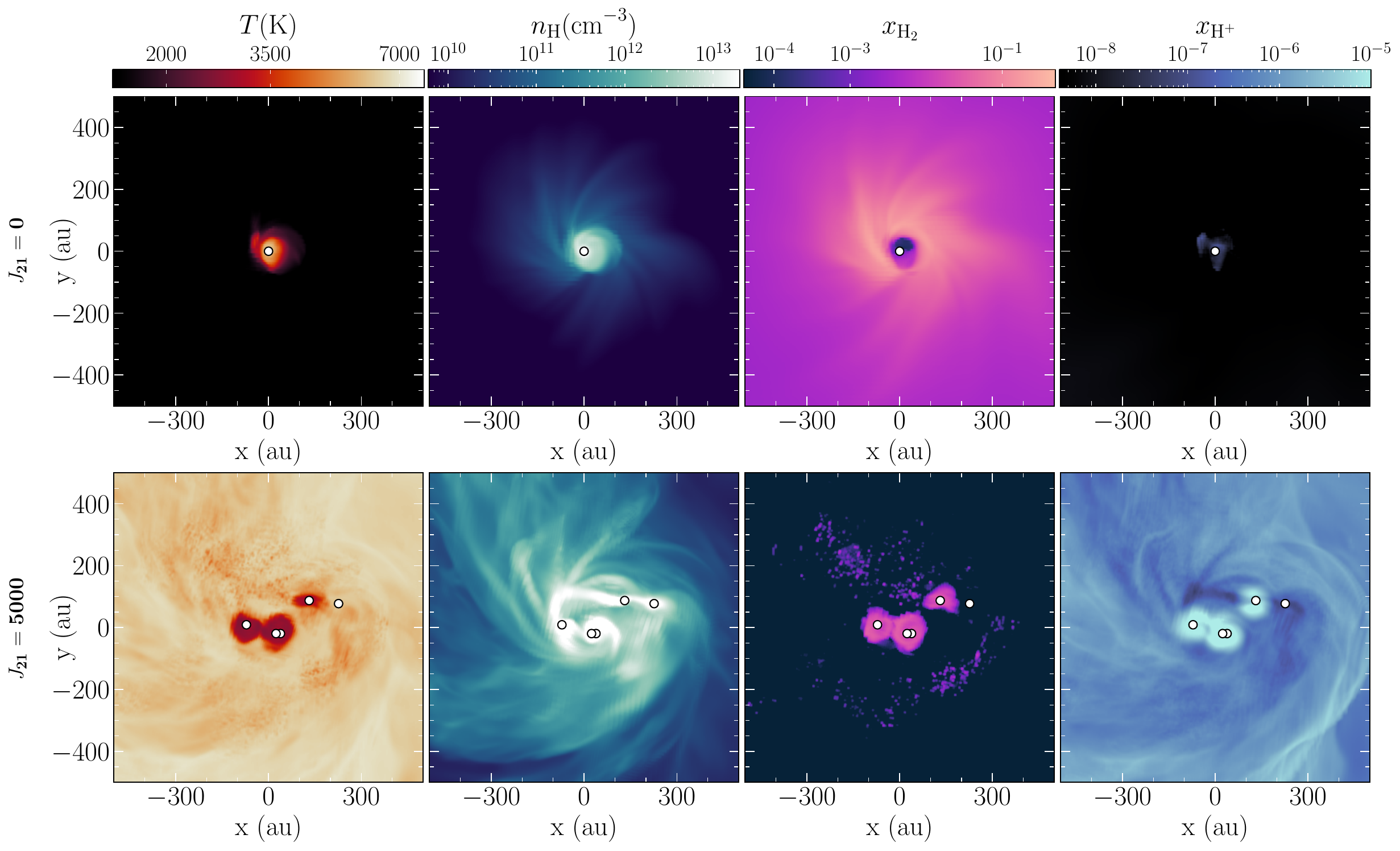}
\caption{Same as \autoref{fig:nobg_overview_basic}, but now comparing the simulations at $z=6$ without (top row) and with (bottom row) external radiation, $200\,\rm{yr}$ after the formation of the first star. Note that the extent of the projection, and the color scales for temperature and number density differ from those in \autoref{fig:nobg_overview_basic}.}
\label{fig:bg_compare_basic}
\end{figure*}

\subsection{What causes the differences in fragmentation at $z=30$ and $z=6$?}
\label{s:results_fragmentation}
We have seen in the previous sections that contrary to expectations, we find less fragmentation in the $z=6$ simulation, despite the fact that the growth of the most massive star is almost identical at $z=6$ and $z=30$. To investigate the underlying reason, we study the evolution of the simulations when a second star appears in the $z=30$ run but not in the $z=6$ run. In \autoref{fig:sink_criteria_nobg}, we look at key criteria that should be fulfilled to create a new sink particle, namely, the grid element where a sink particle is to be created is at the highest level of refinement, shows locally convergent gas motions, is Jeans unstable, is at the (global or local) minimum of the gravitational potential, and is outside the accretion radius of an existing sink particle.

At $z=6$, strong mass inflows ($\nabla \cdot \vec{v} < 0 $) are concentrated near the center at $z=6$, while such inflows also extend to larger radii at $z=30$. Additionally, the location where the second sink appears at $z=30$ has a local gravitational potential minimum, is dense enough to be at the highest level of refinement, and is outside the accretion zone of the existing sink particle. Evidently, this results in the second star formation at $z=30$. At $z=6$, the gravitational potential is limited to one relatively strong, global well, and is steeper than the central potential well at $z=30$. At $z=6$, lower $T_{\rm{CMB}}$ leads to a very dense, central region where all gas is funneled into, and gets accreted onto the existing star without fragmenting to produce another star. The plasma $\beta$ in the outer parts of the disk at $z=6$ is relatively low in comparison to that at $z=30$. Therefore, magnetic support in the disk could also contribute to restricting the conditions for sink formation at $z=6$.

Although we find less fragmentation at $z=6$ in this carefully controlled numerical experiment, it is difficult to generalize this conclusion without running an ensemble of simulations to overcome stochasticity due to turbulence \citep[e.g.,][]{2020MNRAS.494.1871W,Sharda_2020,2023MNRAS.518.4895R}. Nonetheless, this exercise provides meaningful insights into how fragmentation behavior and consequently the Pop III stellar masses and multiplicity can vastly differ across redshifts. It also implies that the Population III IMF varies as a function of cosmic time and environment, and is very sensitive to the interplay between magnetic fields, radiation feedback, and chemistry of the metal-free gas.

\section{Effects of strong LW radiation background at $z=6$}
\label{s:bkg}
The simulations we have discussed so far were run without an external radiation background, which is reasonable at $z = 30$ when no galaxies have yet emerged and the IGM is opaque, but only provides a conservative lower bound on the radiation field strength impinging a pristine star formation site during the EoR. A strong LW background radiation field can dissociate H$_2$, raise the minimum cloud mass required to initiate collapse, and significantly alter primordial star formation \citep[e.g.,][]{1997ApJ...476..458H,2002ApJ...569..558O,2015MNRAS.448..568H,2020MNRAS.492.4386S,2021MNRAS.507.1775S,2021ApJ...917...40K}. Only a handful of studies exist that have explored the effects of LW radiation on primordial star formation by resolving down to $\rm{au}$ scales \citep[e.g.,][]{2008ApJ...673...14O,2023MNRAS.520.2081P}, but they are limited to moderate or weak LW radiation backgrounds appropriate for $z \sim 20$, and do not simultaneously include magnetic fields and radiation feedback. To provide an upper bound on the effects of a background (LW) radiation field on Pop III star formation, we run another RMHD simulation at $z=6$ where we keep all initial conditions identical to the run above,\footnote{One might wonder whether it is appropriate to assume that the small-scale dynamo operates when the LW background is strong ($J_{21} \gg 1$). Recent cosmological simulations by \citet{2024A&A...684A.195D} have shown that this is indeed the case: the small-scale dynamo remains active and amplifies the field even for $J_{21} > 5000$. Thus, it is reasonable to adopt the same initial magnetic field strength in both the simulations.} but add an external LW field $J_{21} = 5000$, where $J_{21}$ is the strength of the field normalized to $10^{-21}\,\rm{erg\,s^{-1}\,cm^{-2}\,Hz^{-1}\,sr^{-1}}$; such high values of $J_{21}$ are rare but can be realized near starburst galaxies during the EoR owing to high escape fractions \citep[e.g.,][]{2000ApJ...534...11H,2008MNRAS.391.1961D,ahn_2009_j21,2020ApJ...897...95V}. Further, such values of $J_{21}$ could also arise from young massive star clusters in the vicinity of metal-free star-forming clumps, which are likely to have high escape fractions \citep[][and references therein]{2025ApJ...987...12M}. To preserve brevity in the text, we will categorize the $z=6$ simulation without the external background as $J_{21} = 0$ while comparing it with the $J_{21} = 5000$ simulation.

As expected \citep[e.g.,][]{2007ApJ...671.1559W,2008ApJ...673...14O}, adding a strong LW radiation background leads a delay in the onset of the collapse, by $\sim$$ 7\,\rm{Myr}$, indicating that gravity must overcome a stronger thermal pressure due to strongly impaired cooling of the gas as LW radiation dissociates $\rm{H}_2$. The initial turbulent energy decays by up to 38 per cent during this period, but this decay is short-lived due to gravity-driven turbulence post the onset of collapse. Once gravity is able to overcome other forces as more gas is funneled towards the cloud core from larger scales, it triggers a runaway cloud collapse, forming multiple protostars that accrete much more rapidly and are very massive as compared to the simulation at $z=6$ with $J_{21} = 0$. Out of the seven stars that form, four have masses $M_{\star} = 173,\,105,\,121$, and $31\,\ \rm{M}_\odot$ within the first $200\,\rm{yr}$, with accretion rates as high as $1\,\rm{M_{\odot}\,yr^{-1}}$. The overall star formation efficiency is almost $5\times$ higher than the $J_{21}=0$ run. When compared to simple power-law accretion profiles, we find that the mass growth follows a more extreme curve than what is seen in the $J_{21}=0$ simulation. Stars follow $M_\star \propto t^p$ with $p \approx 1.0$ in the $J_{21}=5000$ simulation, as compared to $p \approx 0.5$ in the $J_{21}=0$ simulation \citep[][figure 13]{Sharda_2025_popsicle_explanation_notthebigonewithme}. At this rate, they would reach $\sim1000\,\rm{M}_\odot$ within 1000 years, and could act as seeds of intermediate mass or supermassive black holes in the early Universe \citep[e.g.,][]{kelvin_helmholtz_hosokawa_2012,2016PASA...33...51L,2020ARA&A..58...27I,2024MNRAS.534..290L}. Below, we will explore the physical and chemical properties of the externally irradiated star-forming core that gives rise to such vigorous accretion and formation of very massive stars.

\subsection{Chemistry and thermodynamics}
\label{s:bkg_chem}
\autoref{fig:bg_compare_basic} provides a detailed look into the thermodynamic state of the gas for the two $z=6$ simulations with and without an external background radiation. The presence of the background radiation leads to much hotter gas in the core in the $J_{21}=5000$ run as compared to the $J_{21}=0$ run. Nevertheless, the gas column density in regions near the protostars is large enough that H$_2$ shielding allows the gas to be cooler. It is important to note that cross shielding of H$_2$ by H, which has been neglected in previous works, plays an important role alongside H$_2$ self-shielding in ensuring that H$_2$ does not get dissociated and can cool the gas. Moreover, the radiative outputs from the protostars is low given their extremely high accretion rates, further ensuring there is no ionization and dissociation of H$_2$ due to EUV and FUV photons from the protostars in the $J_{21}=5000$ run.

\begin{figure}
\centering
\includegraphics[width=\columnwidth]{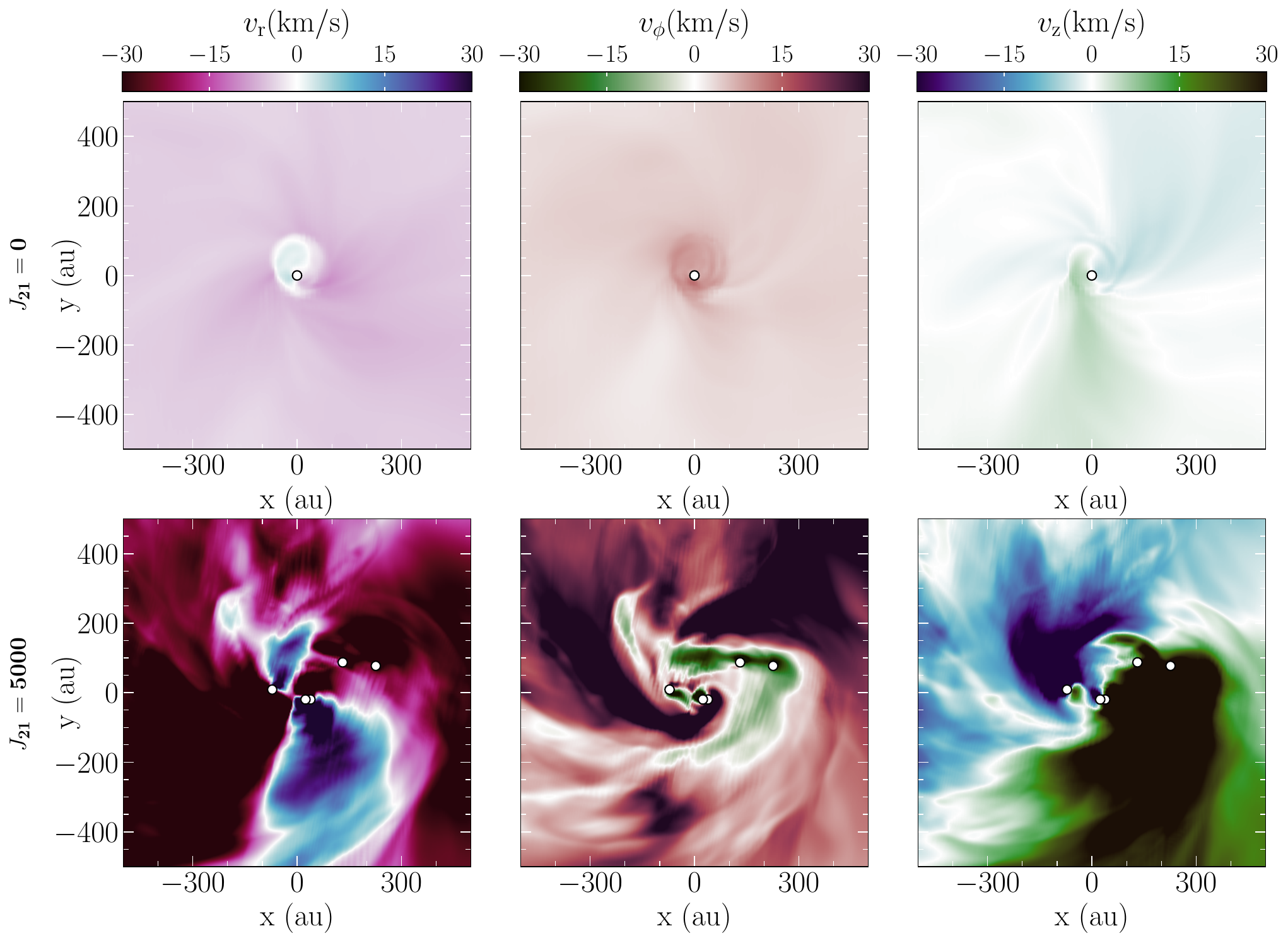}
\caption{Same as \autoref{fig:velocities_nobg}, but for the $z=6$ simulations without (top row) and with (bottom row) external radiation $200\,\rm{yr}$ after the formation of the first star. Note that the extent of the projection, and the color scales for temperature and number density differ from those in \autoref{fig:velocities_nobg}.}
\label{fig:bg_cylvel}
\end{figure}

\begin{figure}
\centering
\includegraphics[width=\columnwidth]{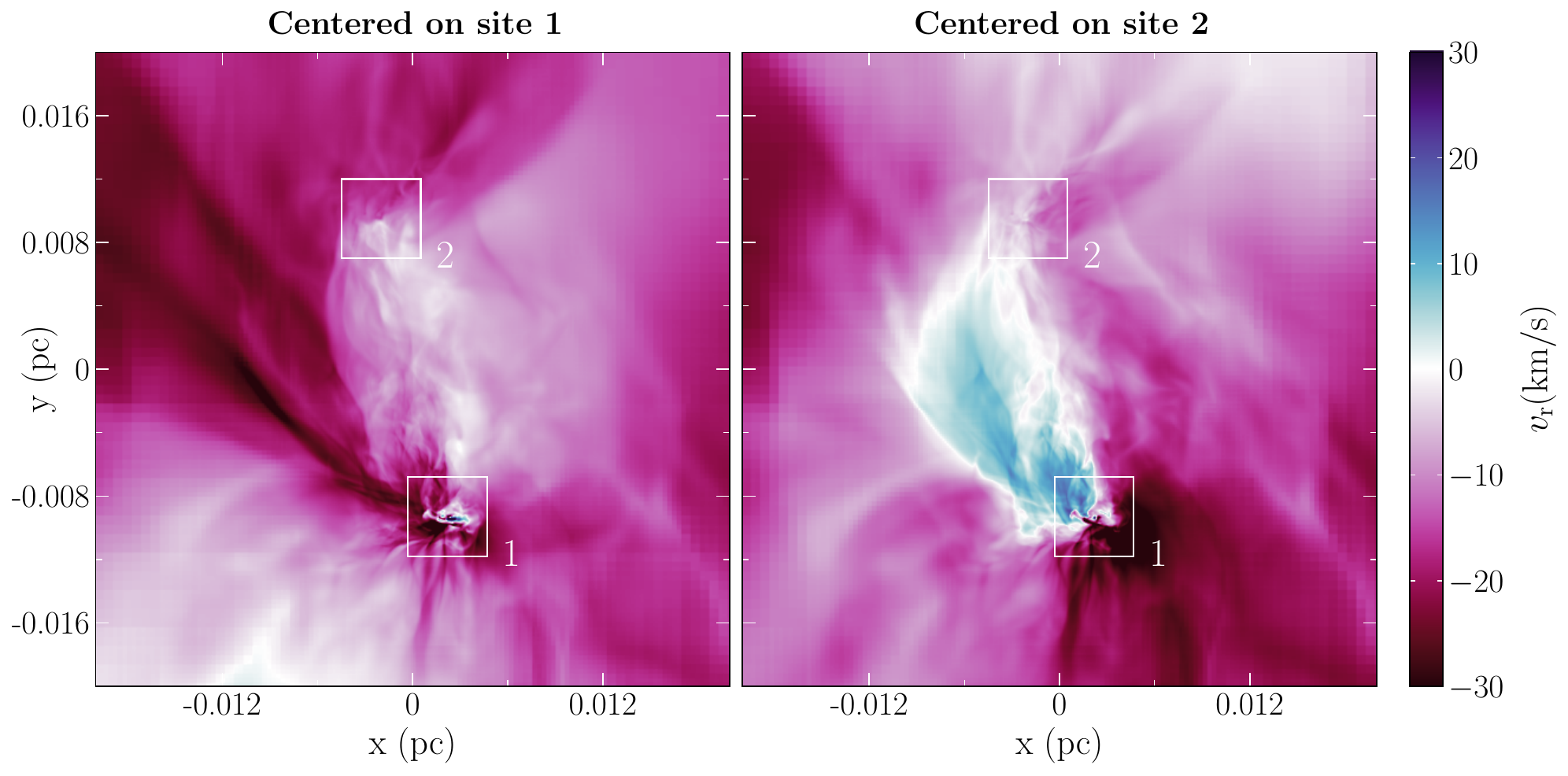}
\caption{Density-weighted radial velocity of the gas, $v_{\rm{r}}$, in the frame of reference of the two star formation sites separated by $\sim 4100\,\rm{au}$, in the $z=6$ simulation with external radiation background. Both the sites are fed by intense gas inflows, leading to high protostellar accretion rates.}
\label{fig:bg_cylvel22}
\end{figure}

We also find that fragmentation is enhanced in the $J_{21}=5000$ simulation, leading to the formation of seven stars within the first $200\,\rm{yr}$; on the contrary, only one star forms in the $J_{21}=0$ run by this point in time. This shows that once collapse is triggered and gas is shielded, formation of multiple protostars is quite likely even in high UV environments \citep[][]{2020MNRAS.492.4386S}. Not all the seven stars are visible in \autoref{fig:bg_compare_basic}, because the two stars form at a site $\sim 4100\,\rm{au}$ away from the central cluster, but are still bound to the system. The two sites are connected by high-density streams that feed the protostars, leading to high accretion rates. It is likely that some of the protostars located at either site will merge, an aspect we explore below in \autoref{s:bkg_merge}.

\begin{figure*}
\centering
\includegraphics[width=\textwidth]{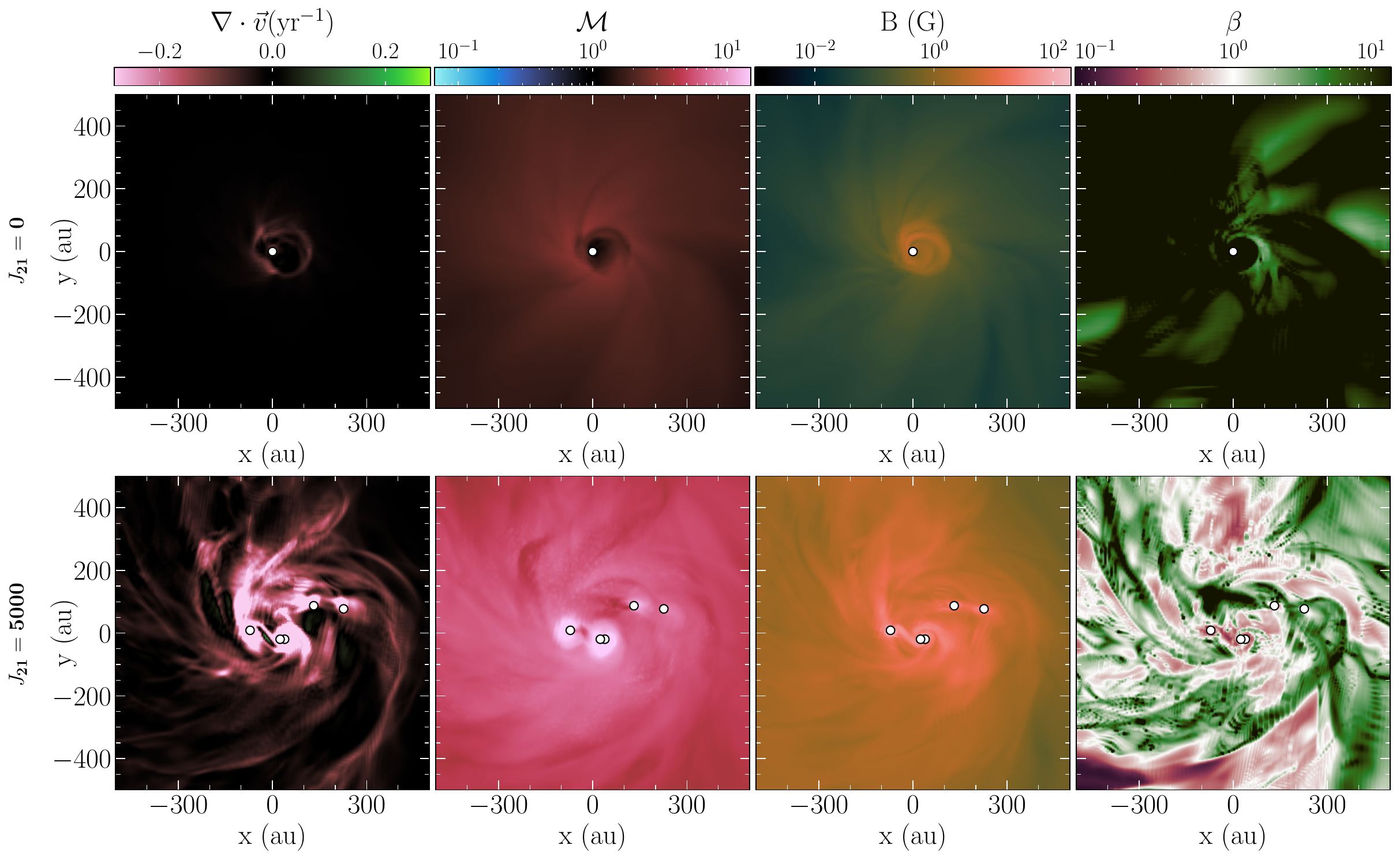}
\caption{Same as \autoref{fig:nobg_mag}, but for the $z=6$ simulations without (top row) and with (bottom row) external radiation $200\,\rm{yr}$ after the formation of the first star. Note that the extent of the projection, and the color scales for temperature and number density differ from those in \autoref{fig:nobg_mag}.}
\label{fig:bg_compare_mag_turb}
\end{figure*}

\subsection{Gas kinematics, magnetic fields and turbulence}
\label{s:bkg_kinematics}
The impact of the LW background on gas kinematics is quite significant. \autoref{fig:bg_cylvel} shows that without a LW background, gas velocities are low, with coherent rotation ($v_\phi > 0$), indicative of the presence of a rotating accretion disk around the protostar. In contrast, runaway collapse in the LW background run induces highly chaotic motions, with extreme radial and azimuthal velocities ($v_{\rm{r,\phi}} \geq 40\,\rm{km\,s^{-1}}$) without coherent rotation. This rapid infall of gas towards the center gives rise to high accretion rates and mass growth we describe above. Zooming out, \autoref{fig:bg_cylvel22} reveals that both the star formation sites in the $J_{21}=5000$ simulation are embedded within large-scale, filamentary gas inflows. The radial velocity structure indicates that each site is fed by intense ($v_{\rm{r}} \sim 30\,\rm{km\,s^{-1}}$), asymmetric accretion streams. These inflows suggest that clustered star formation in the LW-irradiated environment is driven by global gravitational collapse rather than local disk instabilities.

\begin{figure}
\centering
\includegraphics[width=\columnwidth]{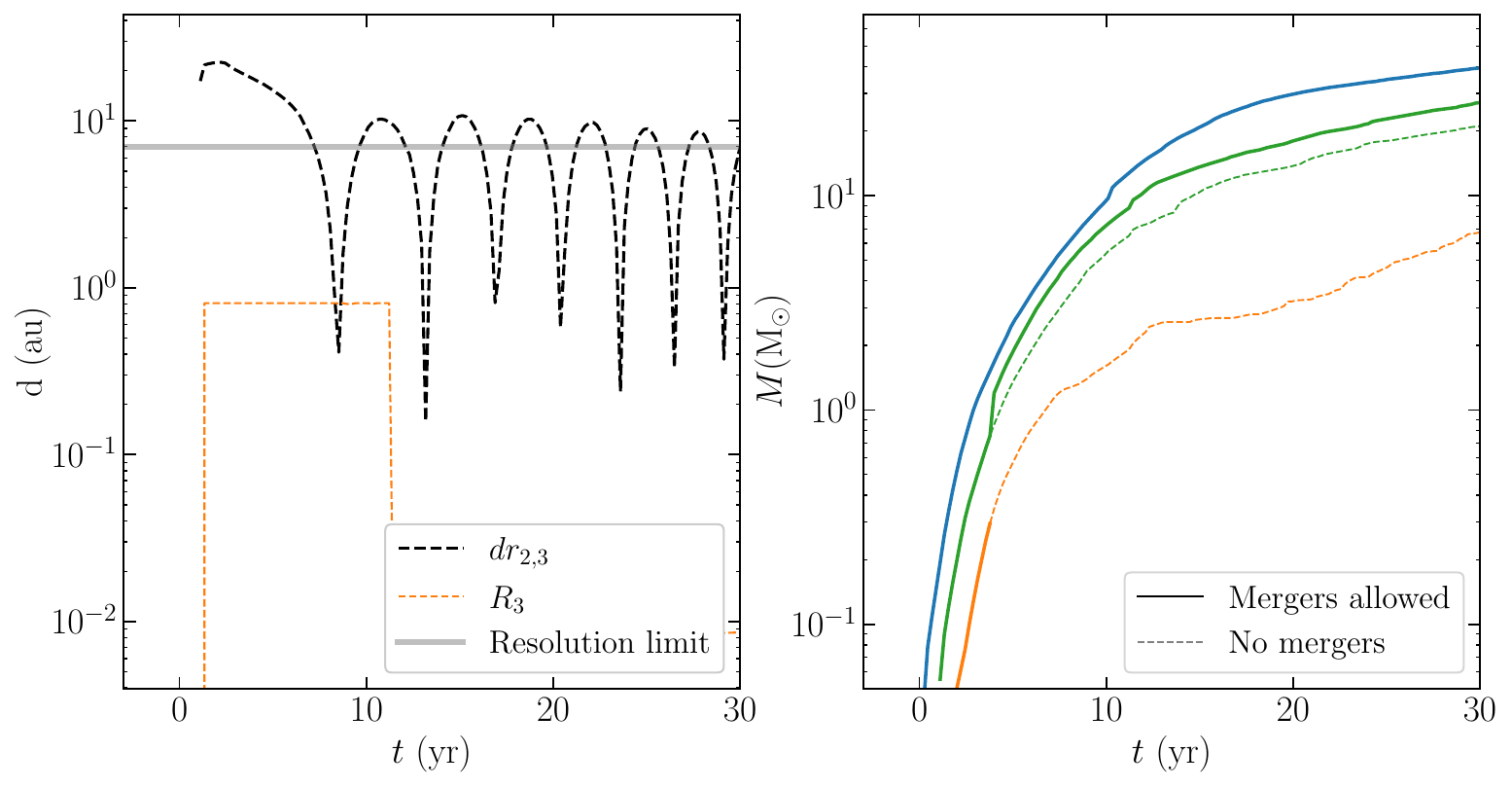}
\caption{\textit{Left panel:} Comparison between the radius of the third sink, $R_3$, and the spatial separation of the second and third sink particles, $dr_{2,3}$ in the $J_{21}=5000$ simulation where sink particle mergers are not allowed. The radius of the second sink is similar to $R_3$, so it was left out for simplicity. The solid gray line shows the resolution limit of the simulation. \textit{Right panel:} Evolution of the protostellar mass in runs with (solid) and without (dashed) sink particle mergers. The epoch of the merger corresponds to the time when the solid and dashed lines diverge. The mass of the third star (orange) is added to that of the second star (green), creating a sudden jump in the stellar mass.}
\label{fig:mergersep}
\end{figure}

\autoref{fig:bg_compare_mag_turb} reveals the turbulent nature of the infalling gas in the $J_{21}=5000$ run that creates intense shocks (negative values in the divergence of the velocity field that are $10\times$ larger than in the $J_{21}=0$ simulations), leading to vigorous accretion onto the newly formed stars. These regions of strong mass inflows also exhibit enhanced magnetic field strengths and highly supersonic Mach numbers. The peak magnetic field strengths reach values of $\geq 10^2\,\rm{G}$ in the $J_{21}=5000$ simulation, which is at least $10\times$ larger than peak strengths we see at the end of the $J_{21} = 0$ run (cf. \autoref{fig:nobg_mag}). The last column of \autoref{fig:bg_compare_mag_turb} shows that several patches across the core have plasma $\beta < 1$ in the $J_{21} = 5000$ run, whereas thermal pressure dominates throughout the core in the $J_{21}=0$ run. Thus, under a strong LW-background, runaway collapse generates a highly turbulent environment with dynamically stronger magnetic fields \citep[see also,][]{regan_2018_popiii_acc_under_lw}. Nevertheless, the magnetic field is weak as compared to gravitational forces, so it cannot significantly hinder gas heating due to compression, and therefore cannot stop the growth of protostars, as would be expected based on earlier work without background radiation \citep[][]{Turk_2012,mckee_mag,Saad_2022_magfields_kin,sharda2025}.

%---------------------------------------
\subsection{Stellar mergers?}
\label{s:bkg_merge}
We do not allow sink mergers in our simulations, as resolving actual protostellar scales ($\lesssim 7.5,\rm{au}$) would be required to capture their likelihood accurately. Modeling stellar mergers also require resolving very short dynamical timescales, which is typically not achievable or desirable in star formation simulations otherwise the long term growth of the protostars cannot be followed \citep[e.g.,][]{2004ApJ...611..399K,bate_2009_merger,greif_2012_merger,2025arXiv251012203F}. Additionally, conserving angular momentum post sink particle mergers also poses numerical challenges \citep[e.g.,][]{2011IAUS..270..425F}, and exploring them is beyond the scope of this work. However, we find that some stars show clear signs of the possibility of a merger in the $J_{21}=5000$ simulation: the left panel of \autoref{fig:mergersep} shows that soon after the second and third stars form, their stellar radii are larger than their separation for a short time period. To test this possibility, we restart the $J_{21}=5000$ simulation from an epoch right before the formation of the first sink particle, this time allowing for sink particle mergers. We caution, however, that such mergers should be interpreted carefully, as their occurrence and outcome are highly sensitive to the adopted numerical prescriptions and resolution limits.

The right panel of \autoref{fig:mergersep} shows that subsequent evolution indeed leads to a merger between the second and third sink, which can be seen in the sudden boost in the mass of the second star. Within 100 yr of this event, the mass of this sink has surpassed that of the first, most massive star. This experiment illustrates how mergers could play a key role in redistributing mass among protostars and altering the multiplicity of the cluster in strong LW radiation environments. Future simulations with the power to resolve down to stellar scales \citep[e.g.,][]{2022MNRAS.512.1430W} and improved merger prescriptions \citep[e.g.,][]{2023MNRAS.522.5180R} will be required to assess their true impact on the Pop III IMF.

%-----------------------------------------------------------------

\section{Conclusions}
\label{s:conclusions}
In this work, we explore the possibility of Population III star formation during the EoR when pockets of gas within and around galaxy haloes could have remained pristine due to inhomogeneous chemical enrichment \citep[e.g.,][]{2016ApJ...823..140X,Venditti_2023,zier2025}. We carry out high-resolution ($7.5\,\rm{au}$) 3D radiation magnetohydrodynamics simulations of collapsing primordial clouds emulating physical conditions appropriate for $z=6$ and $z=30$ as part of the POPSICLE project \citep[][]{Sharda_2025_popsicle_explanation_notthebigonewithme}. Our goal is to understand how Pop III stars grow and evolve in such conditions, and whether any differences are expected in the Pop III IMF during the EoR as compared to $z \geq 20$. We find that:

\begin{enumerate}
    \item In the limit of no external Lyman-Werner radiation ($J_{21}=0$), the key underlying physics that yields differences in the masses of Pop III stars during the EoR (as compared to $z \geq 20$) is the CMB. Contrary to expectations, we observe less fragmentation at $z=6$ (4 stars) than at $z=30$ (12 stars), with the most massive stars in both cases growing only to $M_{\star} \approx 70\,\rm{M_{\odot}}$ due to a combination of feedback from magnetic fields and protostellar radiation.
    \item In the limit of a strong external radiation field ($J_{21}=5000$, possible near starburst galaxies during the EoR), runaway star formation takes place, leading to the formation of very massive stars ($M_{\star} > 100\,\rm{M_{\odot}}$) with vigorous accretion rates and high star formation efficiencies. Our analysis suggests that protostellar mergers are quite likely in such dense and compact clusters.
\end{enumerate}

Although our findings remain to be statistically confirmed, the fact that merely decreasing $T_{\rm{CMB}}$ leads to such differences in Pop III star formation behavior hints that there is no universal Population III IMF \citep[see also,][]{2024ApJ...962...49B}. Our findings also suggest that external radiation field plays a central role in determining the masses of Pop III stars formed during the EoR: strongly irradiated environments could form very massive Pop III stars regardless of turbulence and magnetic fields, whereas weakly irradiated environments could lead to less massive and more compact clusters with larger median stellar masses than at $z \geq 20$.

\begin{acknowledgements}
We thank the anonymous referee for their comments that helped improve the clarity of this work. We thank Alessandra Venditti for useful discussions on the EoR, Robert Schulz for help with technical issues, and Matthieu Schaller for reading a pre-print of this paper. PS is supported by the Leiden University Oort Fellowship and the International Astronomical Union -- Gruber Foundation (TGF) Fellowship. SV acknowledges support from the European Research Council (ERC) Advanced Grant MOPPEX 833460. SHM is supported by the Flatiron Institute, which is a division of the Simons Foundation. This work was performed using the compute resources from the Academic Leiden Interdisciplinary Cluster Environment (ALICE) provided by Leiden University, as well as the Dutch National Supercomputing Facility SURF via project grants EINF-8292 and EINF-12322 on \texttt{Snellius}. We acknowledge using the following softwares: \texttt{FLASH} \citep{flash}, \texttt{KROME} \citep{grassi_2014}, \texttt{VETTAM} \citep{Menon_2022_VETTAM}, \texttt{yt} \citep{2011ApJS..192....9T}, \texttt{paramesh} \citep{2000CoPhC.126..330M}, Numpy \citep{oliphant2006guide,2020arXiv200610256H}, Scipy \citep{2020NatMe..17..261V}, Matplotlib \citep{Hunter:2007}, and \texttt{cmasher} \citep{2020JOSS....5.2004V}.
\end{acknowledgements}

%bib in alphabetical order
\bibliographystyle{aa} % this is the A&A style
\bibliography{references}

%\bibliography{references}
\end{document}